\documentclass[prl, superscriptaddress, showpacs, floatfix,
twocolumn]{revtex4}

\usepackage[]{graphicx}
\usepackage{amsmath}
\usepackage{amsfonts}

\def\one{{\mathchoice{\rm 1\mskip-4mu l}{\rm 1\mskip-4mu l}{\rm 1\mskip-4.5mu l}{\rm
1\mskip-5mu l}}}
\def\ket#1{| #1 \rangle}

\newtheorem{theorem}{Theorem}

\newtheorem{corollary}[theorem]{Corollary}

\newtheorem{lemma}[theorem]{Lemma}

\date{\today}

\begin{document}

\date{\today}
\title{Quantum Error Correcting Subsystems are Unitarily
Recoverable Subsystems}
\author{David~W.~Kribs}
\affiliation{Department of Mathematics and Statistics, University
of Guelph, Guelph, ON, Canada, N1G 2W1} \affiliation{Institute for
Quantum Computing, University of Waterloo, ON Canada, N2L 3G1}
\author{Robert~W.~Spekkens}
\affiliation{Department of Applied Mathematics and Theoretical
Physics, Cambridge University, Cambridge, UK, CB3 0WA}

\begin{abstract}
We show that every correctable subsystem for an arbitrary noise
operation can be recovered by a unitary operation, where the
notion of recovery is more relaxed than the notion of correction
insofar as it does not protect the subsystem from subsequent
iterations of the noise. We also demonstrate that in the case of
unital noise operations one can identify a subset of all
correctable subsystems
---those that can be corrected by a single unitary operation--- as
the noiseless subsystems for the composition of the noise
operation with its dual. Using the recently developed structure
theory for noiseless subsystems, the identification of such
unitarily correctable subsystems is reduced to an algebraic
exercise.
\end{abstract}

\maketitle


A basic principle in quantum computation, communication and
cryptography is the requirement that if quantum information
encoded in a physical system is preserved during some storage or
transmission process, then the information is represented at every
time on a subsystem of the overall system. This heuristic concept
has been formalized as part of the ``subsystem principle'' for
realizing quantum information \cite{VKL01,Kni06}, and provides a
central motivation for quantum error correction techniques.
Recently a framework for studying correctable subsystems for
arbitrary quantum operations was introduced in \cite{KLP05,KLPL05}
under the moniker \textquotedblleft operator quantum error
correction\textquotedblright . The
approach includes standard quantum error correcting (subspace) codes \cite%
{Sho95a,Ste96a,Got96,BDSW96a,KL97a}, and
decoherence-free/noiseless
subspaces and subsystems \cite%
{PSE96,DG97c,ZR97c,LCW98a,KLV00a,Zan01b,KBLW01a}. Although it does
not yield new codes, this approach enables more efficient
correction procedures \cite{Bac05,Pou05,KlaSar06}.


Intuitively, the subsystem principle implies that if a subsystem
is correctable for a given noise operation, then at the output
stage of the transmission there should be an identifiable copy of
the subsystem within the output Hilbert space. Our first main
result in this paper formalizes this notion and thereby
contributes to the rigorous formulation of the subsystem principle
for quantum information. Specifically, we show that every
correctable subsystem for an arbitrary noise operation only
suffers a change in representation due to that operation. A
unitary operation is therefore sufficient to recover from the
noise, in the sense of restoring the quantum information to the
subsystem in which it was encoded, but due to the fact that
complementary subsystems are not necessarily preserved by the
noise, the system is not necessarily protected from subsequent
iteration of the noise operation. We therefore say that the
subsystem is only recovered, but not corrected, by a unitary
operation.


Quantum error-correcting codes are typically designed with
particular noise models in mind. Given the myriad different
physical implementations for quantum information processing and
communication, and the diverse nature of the noise that may
afflict these, techniques for identifying codes for
\emph{arbitrary} noise models are clearly of interest. To date,
this has only been achieved for \emph{passive} quantum error
correction. It is accomplished by identifying the
decoherence-free/noiseless subsystems for the noise model.  A
recently developed structure theory for passive error correction
\cite{CK06,Kni06} reduces the identification of such codes to an
exercise in matrix algebra. (The algorithm is polynomial in the
dimension of the system Hilbert space and can be implemented with
available computer software.) Our second main result demonstrates
that in certain cases finding active quantum error correcting
codes reduces to the problem of finding passive codes.
Specifically, in the case of a noise operation that is unital but
otherwise arbitrary, we identify a subset of the actively
correctable subsystems, those that can be corrected by a single
unitary operation, as precisely the noiseless subsystems for the
composition of the operation with its dual. Hence the
identification of such actively correctable subsystems may be
accomplished via an application of the passive theory
\cite{CK06,Kni06}.


\strut

Let us first discuss nomenclature. Given a quantum system $S$
represented on a (finite-dimensional) Hilbert space
$\mathcal{H}^{S}\equiv \mathcal{H}$, we say a quantum system $B$
is a \textit{subsystem} of $S$ if there is a representation of $B$
on a Hilbert space $\mathcal{H}^{B}$ such that
\begin{equation}
\mathcal{H}^{S}=(\mathcal{H}^{A}\otimes \mathcal{H}^{B})\oplus
\mathcal{K}, \label{subsystem}
\end{equation}%
where $A$ is also a subsystem of $S$ and $\mathcal{K}=(\mathcal{H}%
^{A}\otimes \mathcal{H}^{B})^{\perp }$. We are interested in cases
$\dim \mathcal{H}^{B}>1$, so at least one qubit can be encoded. We
adopt the convention that $\rho $, $\sigma $ and $\tau $ denote
density operators and a superscript such as $\sigma ^{B}$ refers
to the subsystem on which the operator is defined. The set of
operators on $\mathcal{H}$ is denoted by
$\mathcal{L}(\mathcal{H})$.


Linear maps on $\mathcal{L}(\mathcal{H})$ can be regarded as
operators acting on the space $\mathcal{L}(\mathcal{H})$ with the
Hilbert-Schmidt inner product $(\sigma ,\tau )={\mathrm{Tr}}%
(\sigma ^{\dagger }\tau )$.  These maps are called \textit{%
superoperators} to distinguish them from operators acting on
$\mathcal{H}$. We use the terms \emph{channel} and
\emph{operation} to mean a trace-preserving completely positive
linear map $\mathcal{E}:  \mathcal{L}(\mathcal{H})\rightarrow
\mathcal{L}(\mathcal{H}^\prime)$ between Hilbert spaces
$\mathcal{H}$ and $\mathcal{H}^\prime$. For convenience we shall
assume the noise operation $\mathcal{E}$ maps $\mathcal{H}$ to
$\mathcal{H}$, but our results are easily adapted to the general
setting simply by replacing $\mathcal{H}$ with
$\mathcal{H}^\prime$ at the output stage. Such maps describe time
evolution of open quantum systems (in the Schr\"{o}dinger
picture), and can always be represented in the operator-sum
representation $ \mathcal{E}(\sigma )=\sum_{a}E_{a}\sigma
E_{a}^{\dagger }$ by a set of
\textquotedblleft Kraus\textquotedblright\ operators $\{E_{a}\}$ on $%
\mathcal{H}$.  The composition of two superoperators will be
denoted by
$\mathcal{E}\circ\mathcal{F}(\sigma)=\mathcal{E}(\mathcal{F}(\sigma))$.

The \emph{dual} or \emph{adjoint} map $\mathcal{E}^{\dagger }$ is
the superoperator on $\mathcal{L}(\mathcal{H})$ defined in the
usual way by the
equation ${\mathrm{Tr}}(\mathcal{E}^{\dagger }(\sigma )\tau )={\mathrm{Tr}}%
(\sigma \,\mathcal{E}(\tau ))$. A unitary operation $\mathcal{U}$ satisfies $%
\mathcal{U}^{\dag }\circ \mathcal{U}=\mathcal{U}\circ
\mathcal{U}^{\dagger
}=\mathrm{id}$, where $\mathrm{id}$ is the identity superoperator. This is equivalent to the existence of a unitary $U\in \mathcal{L}(%
\mathcal{H})$ that implements the operation via the equation $\mathcal{U}%
(\sigma )=U\sigma U^{\dagger }$. Given a decomposition of
$\mathcal{H}$ as in Eq.~(\ref{subsystem}), $\mathcal{P}_{AB}$
denotes the superoperator defined by $\mathcal{P}_{AB}(\sigma
)=P_{AB}\sigma P_{AB}$ where $P_{AB}$ is the projector onto the
subspace $\mathcal{H}^{AB}=\mathcal{H}^{A}\otimes
\mathcal{H}^{B},$ and let $\mathrm{id}_{B}$ be the identity superoperator on $%
\mathcal{L}(\mathcal{H}^{B})$.
If we are given maps $%
\mathcal{E}_{A}$, $\mathcal{E}_{B}$ on the subsystems, as a
notational convenience we write $\mathcal{E}_{A}\otimes
\mathcal{E}_{B}$ both as a map on
$\mathcal{L}(\mathcal{H}^{A}\otimes \mathcal{H}^{B})$ and for the
natural extension of the map to $\mathcal{L}(\mathcal{H})$.



The definition from \cite{KLP05,KLPL05} of a correctable subsystem
for a quantum operation $\mathcal{E}$ is as follows.\strut

\strut \textbf{Definition:} $B$ is a \textit{correctable
subsystem} for an operation $\mathcal{E}$ if there is a quantum
operation $\mathcal{R},$ which
we call the \emph{correction operation}, and a quantum operation $\mathcal{F}%
_{A}:\mathcal{L}(\mathcal{H}^{A})\rightarrow
\mathcal{L}(\mathcal{H}^{A})$ such that
\begin{equation}
\forall \sigma ^{A}\,\forall \sigma ^{B}\,\,:\,\,(\mathcal{R}\circ \mathcal{E%
})(\sigma ^{A}\otimes \sigma ^{B})=\mathcal{F}_{A}(\sigma
^{A})\otimes \sigma ^{B},  \label{correctable}
\end{equation}%
or equivalently, such that
\begin{equation}
\mathcal{R}\circ \mathcal{E}\circ
\mathcal{P}_{AB}=\mathcal{F}_{A}\otimes \mathrm{id}_{B}.
\label{correctable2}
\end{equation}

\strut It is worthwhile to define a variant of this notion. First
observe that in general there will be many representations of $B$
as a subsystem of $S$. That is, typically there will exist many
subsystems $C$  and Hilbert spaces $\mathcal{K}^\prime$ such that
\begin{equation}
\mathcal{H}^{S}=(\mathcal{H}^{C}\otimes \mathcal{H}^{B})\oplus
\mathcal{K}^\prime.
\end{equation}%
Of course, $C$ can be distinct from $A$, and can even be
associated with a Hilbert space of different dimension. It is the
consideration of different representations of a subsystem that
leads to the following notion.

\strut \textbf{Definition:} $B$ is a \textit{recoverable
subsystem} for an operation $\mathcal{E}$ if there is a quantum
operation $\mathcal{R},$ which
we call the \emph{recovery operation}, and a quantum operation $\mathcal{F}%
_{C|A}:\mathcal{L}(\mathcal{H}^{A})\rightarrow
\mathcal{L}(\mathcal{H}^{C})$ where $C$ is a subsystem of $S$ that
can be different from $A$ such that
\begin{equation}\label{recoverable}
\forall \sigma ^{A}\,\forall \sigma ^{B}\,\,:\,\,(\mathcal{R}\circ \mathcal{E%
})(\sigma ^{A}\otimes \sigma ^{B})=\mathcal{F}_{C|A}(\sigma
^{A})\otimes \sigma ^{B},
\end{equation}%
or equivalently, such that
\begin{equation}\label{recoverable2}
\mathcal{R}\circ \mathcal{E}\circ
\mathcal{P}_{AB}=\mathcal{F}_{C|A}\otimes \mathrm{id}_{B}.
\end{equation}

Note that a correctable subsystem \strut is a special case of a
recoverable subsystem where the subsystem $C$ is equivalent to
$A.$ Furthermore, if $B$ is a recoverable subsystem, then one can
always compose the recovery operation $\mathcal{R}$ with an
additional operation $\mathcal{R}^{\prime }$ that maps all states
of subsystem $C$ to states of $A$ such that the composition
$\mathcal{R}^{\prime }\circ \mathcal{R}$ satisfies the definition
of a correction operation. Consequently, $B$ is a recoverable
subsystem if and only if it is a correctable subsystem. (It is
worth noting that the operation $\mathcal{R}^{\prime }$ will
typically correspond to ``cooling'' the system if the dimension of
C is greater than that of A, however this can always be achieved
by appending to the system an ancilla of dimension at most the
square of that of the system and implementing a unitary operation
on the composite \cite{NC00}.)

There is a significant difference, however, between implementing a
correction operation and implementing a recovery operation. \ In
both cases, quantum information stored in subsystem $B$ can be
brought back to $B$ after a single iteration of the noise
operation $\mathcal{E},$ but only if one implements a
\emph{correction} operation will $B$ be recoverable from a
\emph{subsequent} iteration of the noise operation $\mathcal{E}.$
\ The reason is that if one only implements a recovery operation,
then the state is left in $CB$, while quantum information in $B$
is only guaranteed to be recoverable if the initial state is in
$AB.$

Furthermore, when we constrain the nature of the recovery and
correction operations, the two notions no longer coincide. \
Specifically, we can introduce the following two notions.

\textbf{Definition:} $B$ is a \textit{unitarily correctable
subsystem} for an operation $\mathcal{E}$ if it satisfies the
definition of a correctable subsystem and the correction operation
$\mathcal{R}$ can be chosen to be a
unitary $\mathcal{U},$ so that%
\begin{equation}
\mathcal{U}\circ \mathcal{E}\circ
\mathcal{P}_{AB}=\mathcal{F}_{A}\otimes \mathrm{id}_{B}.
\label{ucorrectable2}
\end{equation}

\strut \textbf{Definition:} $B$ is a \textit{unitarily recoverable
subsystem} for an operation $\mathcal{E}$ if it satisfies the
definition of a recoverable subsystem and the recovery operation
$\mathcal{R}$ can be chosen
to be a unitary $\mathcal{U},$ so that%
\begin{equation}
\mathcal{U}\circ \mathcal{E}\circ
\mathcal{P}_{AB}=\mathcal{F}_{C|A}\otimes \mathrm{id}_{B}.
\label{urecoverable2}
\end{equation}%
\

Note that by operating with $\mathcal{U}^{\dag}$ from the left on
Eqs.~(\ref{ucorrectable2}) and (\ref{urecoverable2}), we obtain an
expression for the action of the noise operation on $AB$.
Heuristically, being unitarily correctable or recoverable ensures
that although quantum information stored in the subsystem $B$ may
be moved to another subsystem, it remains coherent. In the general
case of possibly distinct input and output Hilbert spaces, we note
that the relevant notion to consider is that of {\it isometric}
recovery.

Although a unitarily correctable subsystem is unitarily
recoverable, the converse is not true unless $C$ satisfies $\dim
C=\dim A$ because only in this case can one find a \emph{unitary}
operation $\mathcal{R}^{\prime }$ that maps all states of
subsystem $C$ to states of $A$. \ Thus all we can say in this case
is that a unitarily correctable subsystem is equivalent to a
unitarily recoverable subsystem for which $\dim C$ is at most
$\dim A.$


One might expect that not every correctable subsystem is unitarily
correctable, and this is indeed the case. \ However, every
correctable subsystem is unitarily \emph{recoverable}. \ The
converse of this implication is also true, and together these
facts constitute our first main result.


\begin{theorem}
\label{QECUNS} Let $\mathcal{E}$ be a quantum operation. The
following conditions are equivalent:

(i) $B$ is a correctable subsystem for $\mathcal{E}$.

(ii) $B$ is a unitarily recoverable subsystem for $\mathcal{E}$.
\end{theorem}

The proof is presented in Appendix A. Note that it is a
constructive proof in the sense that if an error model
$\mathcal{E}$ and a correctable
subsystem $B$ are known, then the unitary operation which yields Eq.~(\ref%
{urecoverable2}) may be explicitly obtained. Moreover, as
discussed above, a correction operation can be obtained simply by
composing $\mathcal{U}$ with an additional operation
$\mathcal{R}^\prime$ that maps states of $C$ to states of $A$ such
that $\mathcal{R} = \mathcal{R}^\prime\circ\mathcal{U}$ satisfies
Eq.~(\ref{correctable2}).

There are two special cases of this theorem that are worthy of
note. First, consider the case where $B$ is a correctable
subsystem for $\mathcal{E}$ and $\mathcal{H}^{A}$ is
one-dimensional, so that $\mathcal{H}^{A}\otimes \mathcal{H}^{B}$
is a correctable \emph{subspace} for $\mathcal{E}$.
In this case, $\mathcal{F}_{C|A}$ is simply a density operator
$\omega _{C}$, and Eq.~(\ref{urecoverable2}) becomes $\forall
\sigma ^{B}\,\,:\,\,\mathcal{U}\circ \mathcal{E}(\sigma
^{B})=\omega ^{C}\otimes \sigma ^{B}$. \ Theorem~\ref{QECUNS}
therefore establishes that a correctable subspace for
$\mathcal{E}$ is mapped by $\mathcal{E}$ to a subsystem, modulo an
overall unitary.  This special case was proven by Nayak and Sen
\cite{NaySen06}. That the subspace gets mapped to a subsystem may
be viewed from our general perspective as simply a representation
change of the subsystem $B$.

The second special case of the theorem established previously is
the case $\mathcal{H}=\mathcal{H}^{A}\otimes \mathcal{H}^{B}$, or
equivalently $\mathcal{K}=\{0\}$ in Eq.~(\ref{subsystem}). In
other words, $B$ is associated with a \emph{factor space} of
$\mathcal{H}$ rather than a factor space of a subspace of
$\mathcal{H}$. In this case, any subsystem $C$ in Eq.
(\ref{urecoverable2}) must have the same dimension as the
subsystem $A.$ Consequently, there is no distinction between
unitarily recoverable and unitarily correctable in this case.
Dropping the projection operation $\mathcal{P}_{AB},$ which is the
identity operator in this case,
Eq. (\ref{urecoverable2}) becomes $\mathcal{U}\circ \mathcal{E}=\mathcal{F}%
_{A}\otimes \mathrm{id}_{B}$ or equivalently,
$\mathcal{E}=\mathcal{U}^{\dag
}\circ \left( \mathcal{F}_{A}\otimes \mathrm{id}_{B}\right) .$ Theorem~\ref%
{QECUNS} therefore establishes that if a factor space $B$ is a
correctable subsystem for $\mathcal{E},$ then $\mathcal{E}$ simply
maps $B$ unitarily to
another factor space. This case was proven by Nielsen and Poulin \cite%
{NP05}.

Theorem \ref{QECUNS} allows us to immediately make a nice
connection with the testable
conditions for operator quantum error correction. Taking the adjoint of Eq.~(%
\ref{urecoverable2}), we have
\begin{equation}
\mathcal{P}_{AB}\circ \mathcal{E}^{\dag }\circ \mathcal{U}^{\dag }=\mathcal{F}%
_{C|A}^{\dag }\otimes \mathrm{id}_{B}.  \label{prop2adjoint}
\end{equation}%
Composing Eqs.~(\ref{ucorrectable2},\ref{prop2adjoint}), noting that $%
\mathcal{U}^{\dag }\circ \mathcal{U}=\mathrm{id}$ and defining the
map $\mathcal{G}_{A}$: $\mathcal{B}(\mathcal{H} _{A})\rightarrow
\mathcal{B}(\mathcal{H}_{A})$ by $\mathcal{G}_{A}\equiv
\mathcal{F}_{C|A}^{\dag }\circ \mathcal{F}_{C|A},$ we obtain

\begin{quote}
Testable condition for $B$ to be a correctable subsystem for
$\mathcal{E}:$
There exists a positive superoperator $\mathcal{G}_{A}$ on $A$ such that%
\begin{equation}
\mathcal{P}_{AB}\circ \mathcal{E}^{\dag }\circ \mathcal{E}\circ \mathcal{P}%
_{AB}=\mathcal{G}_{A}\otimes \mathrm{id}_{B}.
\label{superoptestable}
\end{equation}
\end{quote}

\strut This is simply the superoperator form of the standard
testable condition for operator quantum error correction
\cite{KLP05,KLPL05,NP05}. This is seen by noting that two
completely positive maps are equal if and only if the Kraus
operators of one are related by a unitary remixing of the Kraus
operators of the other \cite{NC00}. \ If the Kraus operators of
$\mathcal{E}
$ are denoted by $\{E_{a}\}$ and those of $\mathcal{G}_{A}$ are denoted $%
\{G_{c}\},$ then the Kraus operators of $\mathcal{P}_{AB}\circ \mathcal{E%
}^{\dag }\circ \mathcal{E}\circ \mathcal{P}_{AB}$ are
$\{P_{AB}E_{b}^{\dag }E_{a}P_{AB}\}_{a,b}$ and those of
$\mathcal{G}_{A}\otimes \mathrm{id}_{B}$ are $\{G_{c}\otimes
I_{B}\}_{c},$ so that Eq.~(\ref{superoptestable}) is equivalent to
\begin{equation}
P_{AB}E_{b}^{\dag }E_{a}P_{AB}=F_{ab}\otimes I_{B}
\end{equation}%
where $F_{\alpha }\equiv \sum_{c}u_{\alpha ,c}G_{c}$ is a unitary
remixing of the $G_{c}$. \ This is the testable condition
discovered in \cite{KLP05,KLPL05}. \ It
follows from the results of \cite{NP05} that not only can one derive Eq.~(\ref%
{superoptestable}) from the fact that $B$ is a correctable subsystem for $%
\mathcal{E}$, as we have done, but the opposite implication holds
as well. (It would be interesting to have a version of this
direction of the implication that is native to the superoperator
formalism.)

If $A$ is one-dimensional, $B$ is a subspace of $S$ and the
positive superoperator $\mathcal{G}_{A}$ is simply a positive
scalar $\gamma^{2},$ so that we have
$\mathcal{P}_{B}\circ\mathcal{E}^{\dag}\circ\mathcal{E}\circ
\mathcal{P}_{B}=\gamma^{2}\mathcal{P}_{B}.$ \ This is the
superoperator form of the testable condition for correctable
\emph{subspaces}. The condition was presented in this form in
\cite{NCSB97}.

Finding correctable subsystems (or subspaces) for arbitrary noise
models in full generality appears to be an intractable problem.
Nevertheless, it has been possible to do so in several special
cases, and this has been one of the successes of quantum error
correction. The case of passive error correction is one such
instance. As an application of Theorem~\ref{QECUNS} together with
further analysis, we next show how recently developed techniques
for passive error correction \cite{CK06,Kni06} can be extended to
a special case of active error correction. The notion of a
noiseless (or decoherence-free) subsystem is fundamental to
passive error correction and so we begin with its definition.

\strut \textbf{Definition:} $B$ is a \emph{noiseless subsystem }
for an operation $\mathcal{E}$ if there is an operation
$\mathcal{G}_{A}: \mathcal{L}(\mathcal{H}^{A})\rightarrow
\mathcal{L}(\mathcal{H}^{A})$ such that
\begin{equation}
\forall \sigma ^{A}\,\forall \sigma
^{B}\,\,:\,\,\mathcal{E}(\sigma _{A}\otimes \sigma
_{B})=\mathcal{G}_{A}(\sigma _{A})\otimes \sigma _{B}.
\end{equation}%
or equivalently, such that
\begin{equation}
\mathcal{E}\circ \mathcal{P}_{AB}=\mathcal{G}_{A}\otimes
\mathrm{id}_{B}. \label{invariantsubsystem}
\end{equation}

We will now focus on \emph{unital} quantum operations. Unital
operations are the quantum analogue of classical \textquotedblleft
bistochastic\textquotedblright\ maps (defined in Appendix B) and
are ubiquitous in quantum information and computation \cite{BZ06}.
A unital operation $\mathcal{E}$ is one which takes the identity
operator to itself, $\mathcal{E}(I)=I.$ \ The dual of a unital map
is necessarily trace-preserving because $\mathrm{Tr}(\mathcal{E}
^{\dag }(\sigma ))=\mathrm{Tr}(\mathcal{E}(I)\sigma
)=\mathrm{Tr}(\sigma )$.  Thus, if $\mathcal{E}$ is a unital
channel, then $ \mathcal{E}^{\dagger }$ is a candidate for a
correction operation. Moreover, the dual of a trace-preserving map
is necessarily unital because if $ \mathcal{E}$ is
trace-preserving then $\mathrm{Tr}(\mathcal{E}^{\dag }(I)\sigma
)=\mathrm{Tr}(\mathcal{E}(\sigma ))=\mathrm{Tr}(\sigma )$\ for all
$\sigma $ and consequently $ \mathcal{E}^{\dag }(I)=I.$
Thus, if the superoperator $\mathcal{E}$ is both unital and
trace-preserving, then so is $\mathcal{E}^{\dag }$ and
consequently so is $\mathcal{E}^{\dag }\circ \mathcal{E}$.


It is easy to see that if $B$ is a noiseless subsystem for $\mathcal{%
E}^{\dag }\circ \mathcal{E}$, so that $\mathcal{E}^{\dag }\circ \mathcal{E}%
\circ \mathcal{P}_{AB}=\mathcal{G}_{A}\otimes \mathrm{id}_{B},$
then $B$ is a correctable subsystem for $\mathcal{E}$ because
$\mathcal{E}^{\dag }$ is trace-preserving and thus constitutes a
correction operation that satisfies the definition of
Eq.~(\ref{correctable2}). (The correction in this case simply
involves \textquotedblleft reversing\textquotedblright\ the noise
on the system.) However, one can say more than this, namely, that
$B$ is \emph{unitarily} correctable for $\mathcal{E}$.  Together
with its converse, this fact constitutes our second main result.

\begin{theorem}
\label{mainthm3} Let $\mathcal{E}$ be a unital quantum operation.
The following are equivalent:

(i) $B$ is a unitarily correctable subsystem for $\mathcal{E}$.

(ii) $B$ is a noiseless subsystem for $\mathcal{E}^{\dagger }\circ \mathcal{%
E}$.
\end{theorem}

\strut

The proof relies on a number of ancillary results for unital
operations presented in Appendix~B. The following lemma, which
applies to arbitrary positivity-preserving superoperators, does
not rely on these results and hence we prove it here.

\begin{lemma}
\label{invlemma} Let $\mathcal{O}$ be a positivity-preserving
superoperator. Let $P$ be a projector and let ${\mathcal{P}}(\cdot
)=P(\cdot )P$ be the associated projective superoperator. Then
\begin{equation}
{\mathcal{P}}\circ \mathcal{O}\circ {\mathcal{P}}=\mathcal{O}\circ {\mathcal{%
P}}  \label{inv1}
\end{equation}%
if and only if
\begin{equation}
\mathrm{supp}(\mathcal{O}(P))\subseteq \mathrm{supp}(P)\,,
\label{inv2}
\end{equation}
where $\mathrm{supp}(\sigma)$ denotes the support of the operator
$\sigma$.
\end{lemma}

\noindent \textit{Proof.} The superoperator identity
Eq.~(\ref{inv1})
implies $P\mathcal{O}(P)P=\mathcal{O}(P)$ which is equivalent to Eq.~(\ref%
{inv2}). Conversely, suppose Eq.~(\ref{inv2}) holds and let $\rho
$ be an arbitrary density operator. Noting that
$P-\mathcal{P}(\rho )$ is a positive operator, it follows from the
fact that $\mathcal{O}$ is linear and positivity-preserving that
$\mathcal{O}(P)-\mathcal{O}\circ \mathcal{P}(\rho )$ is also
positive and consequently $\mathrm{supp}(\mathcal{O}\circ
\mathcal{P}(\rho ))\subseteq \,\mathrm{supp}(\mathcal{O}(P))$. Given Eq.~(%
\ref{inv2}), we have $\mathrm{supp}(\mathcal{O}\circ
\mathcal{P}(\rho ))\subseteq \,\mathrm{supp}(P).$ Finally, given
that $\rho $ is arbitrary, Eq.~(\ref{inv1}) follows. QED


\strut

The following lemma, the proof of which can be found in
Appendix~B, is also central to the proof of the theorem.

\begin{lemma}
\strut \label{newlemma}Let $\mathcal{E}$ be a unital operation,
and let $B$ be a correctable subsystem for $\mathcal{E}$. The
following are equivalent:

(i) $\mathrm{supp}(\mathcal{E}^{\dag }\circ
\mathcal{E}(P_{AB}))\subseteq  \mathrm{supp}(P_{AB})$

(ii) $\mathrm{rank}(\mathcal{E}(P_{AB}))=\mathrm{rank}(P_{AB})$
\end{lemma}

\strut  We now provide the proof of Theorem~\ref{mainthm3}. \strut

\noindent \textit{Proof of Theorem~\ref{mainthm3}.} We first
establish the implication $(i)\Rightarrow (ii)$.  Given that $B$
is a correctable subsystem, it follows that the testable
condition of Eq.~(\ref{superoptestable}) holds, that is, $\mathcal{P}%
_{AB}\circ \mathcal{E}^{\dag }\circ \mathcal{E}\circ \mathcal{P}_{AB}=%
\mathcal{G}_{A}\otimes \mathrm{id}_{B}$ (and Theorem~\ref{QECUNS}
provides a simple way of seeing this). \ At this stage, it is
clear that if not for the leading $\mathcal{P}_{AB}$ in this
expression, $B$ would satisfy the
definition of a noiseless subsystem for $\mathcal{E}^{\dag }\circ \mathcal{E%
}$ provided in Eq. (\ref{invariantsubsystem}), namely,
$(\mathcal{E}^{\dag
}\circ \mathcal{E})\circ \mathcal{P}_{AB}=\mathcal{G}_{A}\otimes \mathrm{id}%
_{B}$. \ By Lemma \ref{invlemma}, one sees that the leading $%
\mathcal{P}_{AB}$ can be dropped if $\mathrm{supp}(\mathcal{E}^{\dag }\circ \mathcal{E}%
(P_{AB}))\subseteq \mathrm{supp}(P_{AB}),$ so it remains only to
show that the latter condition is satisfied if $B$ is unitarily
correctable.\emph{\ }

Since $B$ is a unitarily correctable subsystem for $\mathcal{E}$,
there exists a unitary operation $\mathcal{U}$ such that
$\mathcal{U}\circ \mathcal{E}\circ
\mathcal{P}_{AB}=\mathcal{F}_{A}\otimes \mathrm{id}_{B},$ from
which one
finds that $\mathcal{E}(P_{AB})=\mathcal{U}^{\dag }(\mathcal{F}%
_{A}(I_{A})\otimes I_{B}).$ \ It follows that $\mathrm{rank}(\mathcal{E}%
(P_{AB}))=\mathrm{rank}(\mathcal{F}_{A}(I_{A}))\mathrm{rank}(I_{B})\leq
\mathrm{rank}(I_{A})\mathrm{rank}(I_{B})=\mathrm{rank}(P_{AB})$,
where we
have made use of the fact that $P_{AB}=I_{A}\otimes I_{B}.$ \ But $\mathcal{E%
}$ is unital, so by Corollary \ref{corollaryofuhlmann} (provided
in Appendix~B), we have $\mathrm{rank}(\mathcal{E}(P_{AB}))\geq
\mathrm{rank}(P_{AB}).$ \ One concludes that there must be
equality between the two quantities, that
is, $\mathrm{rank}(\mathcal{E}(P_{AB}))=\mathrm{rank}(P_{AB}).$ \ By Lemma %
\ref{newlemma}, it follows that \strut
$\mathrm{supp}(\mathcal{E}^{\dag }\circ
\mathcal{E}(P_{AB}))\subseteq \mathrm{supp}(P_{AB}).$

\strut We now prove $(ii)\Rightarrow (i)$. From the fact that $B$
is a noiseless
subsystem, we infer that \textrm{supp}$(\mathcal{E}^{\dag }\circ \mathcal{E}%
(P_{AB}))\subseteq $ \textrm{supp}$(P_{AB}).$ \ By Lemma
\ref{newlemma}, it follows that
$\mathrm{rank}(\mathcal{E}(P_{AB}))=\mathrm{rank}(P_{AB}).$ \ Now,
given that $B$ is a correctable subsystem for $\mathcal{E}$ (as
established in the discussion above Theorem \ref{mainthm3}), by
Theorem~\ref{QECUNS} it is also a unitarily recoverable subsystem
for $\mathcal{E}$, so that there exists a unitary operation
$\mathcal{U}$ such that $\mathcal{U}\circ \mathcal{E}\circ
\mathcal{P}_{AB}=\mathcal{F}_{C|A}\otimes \mathrm{id}_{B}.$ \
Given that $P_{AB}=I_{A}\otimes I_{B},$ it follows that
$\mathcal{U}\circ
\mathcal{E}(P_{AB})=\mathcal{F}_{C|A}(I_{A})\otimes I_{B},$ and
consequently
that $\mathrm{rank(}\mathcal{E}(P_{AB}))=\mathrm{rank(}\mathcal{F}%
_{C|A}(I_{A})) \mathrm{rank(}I_{B})$. But given that $\mathrm{rank}(%
\mathcal{E}(P_{AB}))=\mathrm{rank}(P_{AB})=\mathrm{rank}(I_{A})\mathrm{rank}%
(I_{B}),$ we conclude that $\mathrm{rank(}\mathcal{F}_{C|A}(I_{A}))=\mathrm{%
rank(}I_{A}).$ \ This implies that there exists a unitary
$\mathcal{V}$ such
that $\mathcal{V}\circ (\mathcal{F}_{C|A}\otimes \mathrm{id}_{B})=\mathcal{F}%
_{A}\otimes \mathrm{id}_{B}$ for some operation $\mathcal{F}_{A}$
on
subsystem $A.$ \ Consequently, $\mathcal{V}\circ \mathcal{U}\circ \mathcal{E}%
\circ \mathcal{P}_{AB}=\mathcal{V}\circ \mathcal{F}_{C|A}\otimes \mathrm{id}%
_{B}=\mathcal{F}_{A}\otimes \mathrm{id}_{B},$ which by Eq. (\ref%
{ucorrectable2}) implies that the unitary operation
$\mathcal{V}\circ \mathcal{U}$ is a correction operation for
$\mathcal{E},$ and thus that $B$ is unitarily correctable. QED

\strut

By combining Theorem\ \ref{mainthm3} with the recently developed
structure theory for noiseless subsystems \cite{CK06,Kni06}, we
obtain a method for finding the unitarily correctable subsystems
for any unital operation $\mathcal{E}$.
Specifically, the noiseless subsystems of $\mathcal{E}^{\dag }\circ \mathcal{%
E}$ are obtained from the fixed point set
$\mathrm{Fix}\,(\mathcal{E}^{\dag }\circ \mathcal{E})=\{\sigma
:(\mathcal{E}^{\dag }\circ \mathcal{E})(\sigma
)=\sigma \}$, in the following way. This set is a $\dagger $-algebra \cite%
{Kri03a} and the representation theory for such algebras induces a
Hilbert space decomposition $\mathcal{H}=\oplus
_{k}(\mathcal{H}^{A_{k}}\otimes
\mathcal{H}^{B_{k}})$ in which $B$ is a noiseless subsystem for $\mathcal{E}%
^{\dag }\circ \mathcal{E}$ if and only if $\mathcal{H}^{B}\subseteq \mathcal{%
H}^{B_{k}}$ for some $k$ \cite{CK06}. (See \cite{Kni06,HKL04} for
further discussions and analysis.) The fixed point set of an
operation $\mathcal{E}$
is simply the eigenvalue-1 operator eigenspace of the superoperator $%
\mathcal{E}$, which is straightforward to determine.

For an arbitrary unital noise operation $\mathcal{E},$ not every
correctable subsystem is unitarily correctable, and consequently
the noiseless subsystems of $\mathcal{E}^{\dagger }\circ
\mathcal{E}$ do not in general capture all correctable codes for a
typical unital channel $\mathcal{E}$.

The generic two-qubit \textquotedblleft binary unitary
channels\textquotedblright\ provide a class of operations that
illustrate
this point \cite{CKZ05b}. As an example, let $U$ be a unitary on $\mathcal{H}%
={\mathbb{C}}^{4}={\mathbb{C}}^{2}\otimes{\mathbb{C}}^{2}$ with
distinct eigenvalues $\lambda _{j}=\exp (i\theta _{j})$ ordered so
that $0\leq \theta _{1}<\theta _{2}<\theta _{3}<\theta _{4}<2\pi
$. Let $|\psi _{j}\rangle $, $j=1,2,3,4$, be corresponding
eigenstates. Fix a probability $0<p<1$, and define a unital channel $%
\mathcal{E}$ by $\mathcal{E}(\sigma )=p\sigma +(1-p)U\sigma
U^{\dagger }$.
Then $\mathcal{E}$ has two Kraus operators given by, up to normalization, $%
\mathcal{E}=\{I,U\}$. The noiseless subsystems (actually subspaces
in this case) for $\mathcal{E}^{\dagger }\circ \mathcal{E}$ come
from the so-called noise commutant $\{I^{\dagger }I,I^{\dagger
}U,U^{\dagger }I,U^{\dagger }U\}^{\prime }=\{U,U^{\dagger
}\}^{\prime }=\{U\}^{\prime }$, which coincides with the fixed
point set. But since $U$ has distinct eigenvalues, this commutant
is isomorphic as a $\dagger$-algebra to the algebra
${\mathbb{C}}\oplus{\mathbb{C}}\oplus{\mathbb{C}}\oplus{\mathbb{C}}$,
and hence can only be used to encode classical information.

On the other hand, this channel has correctable qubit codes. For
example,
let $\lambda$ be the point of intersection of the line segments $%
[\lambda_1,\lambda_3]$ and $[\lambda_2,\lambda_4]$. Let $s$ and
$t$ be fixed probabilities such that $\lambda = s \lambda_1 +
(1-s)\lambda_3 = t \lambda_2 + (1-t)\lambda_4$, and define states
$\{\ket{\psi},\ket{\phi}\}$ by
\begin{eqnarray*}
| \psi \rangle &=& \sqrt{s}\, | \psi_1 \rangle + \sqrt{1-s} \,|
\psi_3
\rangle \\
| \phi \rangle &=& \sqrt{t} \,| \psi_2 \rangle + \sqrt{1-t} \,|
\psi_4 \rangle .
\end{eqnarray*}
The two-dimensional subspace spanned by $| \psi \rangle$ and $|
\phi \rangle$ is a correctable code for $\mathcal{E}$. In
particular, one can compute that $PUP=\lambda P$ where $P= | \psi
\rangle\langle \psi | + | \phi \rangle\langle \phi |$, and thus
the error correction condition from \cite{KL97a} is satisfied for
$\mathcal{E}$ on the subspace $\mathcal{C}={\rm
span}\,\{\ket{\psi},\ket{\phi}\}$.

\strut

There are many unital noise operations, however, for which the
composition of this map with its dual \emph{does} have noiseless
subsystems. As a simple example, consider the swap operation $|
\psi \rangle\otimes| \phi \rangle\mapsto| \phi \rangle\otimes|
\psi
\rangle $ on a composite quantum system ${\mathcal{H}}= {\mathcal{H}}%
^A\otimes{\mathcal{R}}^A$ made up of a subsystem ${\mathcal{H}}^A$
and a replication ${\mathcal{R}}^A = {\mathcal{H}}^A$. It is clear
that both the subsystem ${\mathcal{H}}^A$ and its copy can be
returned to their initial locations by simply applying the swap
operation again (which is equal to its dual).

Of course, one could note that the swap operation itself has a
noiseless subsystem of the same size; namely the symmetric space
$| \psi \rangle\otimes| \psi \rangle$. But it is easy to find
examples of operations with no noiseless subsystem, for which the
composition map has a non-trivial noiseless subsystem.

To this end, consider a two-qubit system exposed to decoupled
phase flips. The associated error model satisfies
${\mathcal{E}}(\rho) = p Z_1\rho Z_1 + (1-p) Z_2 \rho Z_2$ for
some fixed probability $0<p<1$ and $Z_1=Z\otimes\one_2$,
$Z_2=\one_2\otimes Z$. In this case there is no noiseless
subsystem (or subspace) for ${\mathcal{E}}$. This follows from the
fact that the noise commutant $\{Z_1,Z_2\}^\prime$ is isomorphic
to the algebra
${\mathbb{C}}\oplus{\mathbb{C}}\oplus{\mathbb{C}}\oplus{\mathbb{C}}$.
Thus, only classical information can be safely sent through the
channel unscathed.

However, the operators supported on the subspace spanned by $| 0_L
\rangle = | 00 \rangle$ and $| 1_L \rangle=| 11 \rangle$ form a
noiseless subspace for $\mathcal{E}^\dagger\circ\mathcal{E}$.
Indeed, the set of operators $\sigma = a |00\rangle\!\langle 00 |
+b |00\rangle\!\langle 11 | +c |11\rangle\!\langle 00 | +d
|11\rangle\!\langle 11 |$ form a subalgebra of the commutant $
\mathrm{Fix}\,(\mathcal{E} ^{\dag }\circ\mathcal{E}) = \{
Z_1^\dagger Z_2, Z_1^\dagger Z_1, Z_2^\dagger Z_1 Z_2^\dagger
Z_2\}^\prime = \{ Z_1Z_2\}^\prime$. The unitary correction
operation guaranteed by Theorem~\ref{mainthm3} in this case
happens to be the controlled phase flip operation $U =
|00\rangle\!\langle 00 | +  |01\rangle\!\langle 01 | +
|10\rangle\!\langle 10 | - |11\rangle\!\langle 11 |$.

\strut

{\noindent }\textit{Conclusion.} --- We showed that every
correctable subsystem for an arbitrary quantum operation is a
unitarily recoverable subsystem for the operation. Thus, the
effect of the operation on the subsystem can be reversed by a
single unitary operation, up to a change in the representation of
the subsystem. Our proof was constructive in nature, showing
explicitly how the representation and unitary may be obtained, and
hence a correction operation, if the operation and correctable
subsystem are known. We also suggested that this result
contributes to the rigorous formulation of the subsystem principle
for quantum operations.

We showed that the unitarily correctable subsystems for unital
quantum operations are precisely the noiseless subsystems for the
operation followed by its dual. We indicated how such subsystems
can be practically computed and discussed some simple examples.
The possibility of extending this result to the case of nonunital
quantum operations remains a problem for future research.

\strut

\noindent \textit{Acknowledgements.} D.W.K. was partially
supported by NSERC, CFI and OIT. R.W.S. acknowledges support from
the Royal Society. We would like to thank Stephen Bartlett, Robin
Blume-Kohout, Raymond Laflamme, Fotini Markopoulou, David Poulin,
and Paolo Zanardi for helpful discussions. \ We also thank Robin
Blume-Kohout for suggesting the distinction between correctable
and recoverable subsystems, which simplified the statement of our
results. His thesis \cite{Blu05} also inspired the superoperator
approach adopted here.


\strut

\section{\strut Appendix A: Proof of Theorem \protect\ref{QECUNS} }

\textit{Proof of Theorem \ref{QECUNS}.} The implication
$(ii)\Rightarrow (i)$ follows immediately from the fact that a
unitarily recoverable subsystem is a special case of a recoverable
subsystem, and as discussed below Eq.~(\ref{recoverable2}) being
recoverable implies being correctable.

For $(i)\Rightarrow (ii)$, let $\mathcal{E} = \{E_a\}$ be a Kraus
operator representation for $\mathcal{E}$ and assume $B$ is
correctable for $\mathcal{E}$.

We proceed as follows. First we shall construct an operation
$\mathcal{G}=\{G_a\}$ such that the operators $\{ G_a P_{AB}\}$
have mutually orthogonal ranges, that is $P_{AB} G_a^\dagger G_b
P_{AB} = 0$ for all $a\neq b$, and $\mathcal{G}(I_A\otimes
\sigma^B) = \mathcal{E}(I_A\otimes\sigma^B)$ for all $\sigma^B$.
Then we shall find a subsystem $C$ such that $CB$ can be
identified with a subspace of $\mathcal{H}$, an operation
$\mathcal{F}_{C|A}:\mathcal{L}(\mathcal{H}^A)\rightarrow\mathcal{L}(\mathcal{H}^C)$,
and a unitary operation $\mathcal{V}$ on
$\mathcal{L}(\mathcal{H})$, such that
\begin{equation}
\mathcal{G}(\sigma ^{A}\otimes \sigma
^{B})=\mathcal{V}(\mathcal{F}_{C|A}(\sigma ^{A})\otimes \sigma
^{B})\quad \forall \sigma ^{A}\,\,\forall \sigma ^{B}.
\end{equation}
Then we will have
\begin{equation}
\mathcal{V}^\dagger\circ\mathcal{E}(I_A\otimes\sigma^B) =
\mathcal{F}_{C|A}(I_A)\otimes \sigma ^{B} \quad \forall \sigma^B,
\end{equation}
and we can use a customary positivity-cum-linearity argument (c.f.
Lemma~2.3 \cite{KLPL05}) to show that Eq.~(\ref{urecoverable2})
holds for all $\sigma^A$.

We begin by noting that the testable conditions from \cite{KLP05,KLPL05,NP05} give us operators $%
F_{ab}$ on $\mathcal{H}^A$ such that
\begin{eqnarray}  \label{Fab}
P_{AB} E_a^\dagger E_b P_{AB} = F_{ab} \otimes I_B \quad \forall
a,b.
\end{eqnarray}
Observe that the operator block matrix $F=(F_{ab})$ is positive since $%
(I_m\otimes P_{AB}) E^\dagger E (I_m\otimes P_{AB}) = F \otimes
I_B$, where the row matrix $E = [E_{a_1} \, E_{a_2} \, \cdots
\,]$, the number of $E_a$ is $m$, and $I_m$ is the identity
operator on $m$-dimensional Hilbert space.

Next let $U$ be a unitary such that $UFU^\dagger = D $ is
diagonal, and let $U=(U_{ab})$ and $D= (D_{ab})$ be the associated
block decompositions. Then
\begin{equation}  \label{unitary1}
\sum_{c,d} U_{ac} F_{cd} U_{bd}^\dagger = \delta_{ab} D_{aa} \quad
\forall a,b,
\end{equation}
\vspace{-0.15in}
\begin{equation}  \label{unitary2}
\sum_c U_{ca}^\dagger U_{cb} = \delta_{ab} I_A \quad \forall a,b.
\end{equation}
Define a superoperator $\mathcal{G}=\{G_a\}$ where for all $a$,
\begin{equation*}
G_a = \sum_b E_b (U_{ab}^\dagger \otimes I_B)P_{AB} \,\,\, + E_a
P_{AB}^\perp.
\end{equation*}
Let $X_{ab} = E_b (U_{ab}^\dagger \otimes I_B)P_{AB}$. Then by Eqs.~(\ref%
{Fab},\ref{unitary1}), one can verify for all $a$, $b$
\begin{eqnarray*}
P_{AB}G_a^\dagger G_b P_{AB} &=& \sum_{c,d} X_{ac}^\dagger X_{bd} \\
&=& \Big(\sum_{c,d} U_{ac}F_{cd}U_{bd}^\dagger\Big)\otimes I_B \\
&=& D_{ab}\otimes I_B,
\end{eqnarray*}
and $D_{ab}=0$ for all $a\neq b$. Moreover, Eq.~(\ref{unitary2})
yields for all $\sigma^B$
\begin{eqnarray*}
\mathcal{G}(I_A \otimes \sigma^B) &=& \sum_a G_a(I_A \otimes
\sigma^B)
G_a^\dagger \\
&=& \sum_{a,b,c} X_{ab} (I_A \otimes \sigma^B) X_{ac}^\dagger \\
&=& \sum_{b,c} E_b (\Big( \sum_a U_{ab}^\dagger U_{ac}\Big)
\otimes \sigma_B
) E_c^\dagger \\
&=& \sum_b E_b (I_A \otimes \sigma^B)E_b^\dagger \\
&=& \mathcal{E}(I_A\otimes \sigma^B).
\end{eqnarray*}

To simplify notation, let $\dim B = n$ and fix an orthonormal
basis $\{| \psi_k \rangle\}_{k=1}^n$ for $B$. Let $m$ be the
cardinality of the set $\{ a : D_{aa}\neq 0\}$, and for all $a$
let $r_a = \mathrm{rank}\,\sqrt{D_{aa}} = \mathrm{rank}\, D_{aa}$.

By the polar decomposition applied to each $G_aP_{AB}$, and the
fact that these operators have mutually orthogonal ranges, there
are partial isometries $V_a$ with mutually orthogonal ranges for
distinct $a$ such that
\begin{equation}
G_a P_{AB} = V_a \sqrt{P_{AB} G_a^\dagger G_a P_{AB}} = V_a (\sqrt{D_{aa}}%
\otimes I_B).
\end{equation}

It follows that we can find a decomposition of each $V_a$ of the
form
\begin{equation}
V_a = \sum_{k=1}^n \sum_{l=1}^{r_a} | \pi_l^{(a)} \rangle|
\theta_k^{(a)} \rangle\langle \psi_k |\langle \phi_l^{(a)} |,
\end{equation}
where $\{| \phi_l^{(a)} \rangle\}_{l=1}^{r_a}$ is an orthonormal basis for $%
\mathrm{range}\,\sqrt{D_{aa}} = \mathrm{range}\,D_{aa}$ for each
$a$. The vectors $| \pi_l^{(a)} \rangle| \theta_k^{(a)} \rangle =
V_a | \phi_l^{(a)}
\rangle| \psi_k \rangle$ are given by the tensor product on  the final space $V_a V_a^\dagger%
\mathcal{H}$ induced by the tensor product on the initial space
$V_a^\dagger V_a\mathcal{H}$ and the isometric action of $V_a:
V_a^\dagger V_a\mathcal{H}\rightarrow V_aV_a^\dagger\mathcal{H}$.

Now let $\{| w_a \rangle\}_{a=1}^m$ be an orthonormal set of vectors in an $%
m$-dimensional Hilbert space. The vectors $\{| \phi_l^{(a)}
\rangle| w_a \rangle| \psi_k \rangle\}_{a,k,l}$ are orthonormal
since the $| w_a \rangle$ are, and because, for a fixed $a$, the
vectors $\{| \phi_l^{(a)} \rangle| \psi_k \rangle\}_{k,l}$ form an
orthonormal set. Moreover, the orthogonality of the ranges of the
operators $G_a P_{AB}$ ensures the dimension of $\mathcal{H}$ is
at least $\sum_{a=1}^m n\,r_a$. Thus, we may identify the Hilbert
space spanned by the vectors $\{| \phi_l^{(a)} \rangle| w_a
\rangle| \psi_k \rangle\}_{a,k,l}$ with a subspace of
$\mathcal{H}$.

Next define $V$ to be any unitary extension of the following
partial isometry to all of $\mathcal{H}$:
\begin{equation}
V \,\, : \,\, | \phi_l^{(a)} \rangle| w_a \rangle| \psi_k \rangle
\longmapsto | \pi_l^{(a)} \rangle| \theta_k^{(a)} \rangle.
\end{equation}
The image vectors $\{| \pi_l^{(a)} \rangle| \theta_k^{(a)}
\rangle\}_{a,k,l}$ form an orthonormal set since the $V_a$ are
partial isometries with mutually orthogonal ranges, and because
these vectors are orthonormal for a fixed $a$. Further define
$\mathcal{F}_{C|A} = \{ \sqrt{D_{aa}} \otimes | w_a \rangle \}$ as
a channel on $\mathcal{L}(\mathcal{H}^A)$, and so for all
$\sigma^A$,
\begin{equation}
\mathcal{F}_{C|A}(\sigma^A) = \sum_a (\sqrt{D_{aa}}\sigma^A\sqrt{D_{aa}}%
)\otimes|w_a\rangle\!\langle w_a |.
\end{equation}
Finally, it can be verified by direct computation that
\begin{equation}
\mathcal{G}(\sigma ^{A}\otimes \sigma
^{B})=\mathcal{V}(\mathcal{F}_{C|A}(\sigma ^{A})\otimes \sigma
^{B})\quad \forall \sigma ^{A}\,\,\forall \sigma ^{B},
\label{unstons}
\end{equation}%
and this completes the proof. QED

\section{Appendix B: Theorem~\ref{mainthm3}}\label{AppB}

 In this appendix we establish results used in the proof
of Theorem~\ref{mainthm3}.

We begin by recalling a result for classical maps. A vector of
probabilities $\mathbf{p}$ of dimension $n$ is said to be
\emph{majorised} by a vector $\mathbf{q}$, denoted
$\mathbf{p}\prec\mathbf{q}$, if for each $k$ in the range 1 to
$n$, $\sum_{j=1}^k p_j^{\downarrow} \le \sum_{j=1}^k
q_j^{\downarrow}$, with equality for $k=n$, where the $\downarrow$
indicates that the probabilities are to be taken in nonincreasing
order. A \emph{bistochastic map} is a matrix $\Lambda$ satisfying
$\sum_{k}\Lambda_{kk^{\prime }}=1$ and
$\sum_{k^{\prime}}\Lambda_{kk^{\prime }}=1$.  The
Hardy-Littlewood-Polya theorem \cite{HLP34,Bha97} states that the
output of a bistochastic map is majorised by the input, that is,
if $\mathbf{p}=\Lambda\mathbf{q}$ for bistochastic $\Lambda$, then
$\mathbf{p}\prec\mathbf{q}$.

A similar result holds for unital operations. \ It is a theorem
due to Uhlmann \cite{Uhl77,BZ06}. We include a short proof for
completeness.

\begin{lemma}
\label{generalizeduhlmann} If $\rho=\mathcal{E}(\sigma)$ for a
unital channel $\mathcal{E}$, then the ordered spectrum
$\mathbf{r}$ of $\rho$ is majorised by the ordered spectrum
$\mathbf{s}$ of $\sigma$; that is,
\begin{equation}\label{uhlmanneqn}
\rho =\mathcal{E}(\sigma)\text{ with }\mathcal{E}\text{ unital implies }%
\mathbf{r}\prec\mathbf{s.}
\end{equation}
\end{lemma}

\noindent\textit{Proof.} Let $p_{k}$ and $\left\vert
e_{k}\right\rangle $
(respectively, $q_{k}$ and $\left\vert f_{k}\right\rangle )$ denote the $k$%
th eigenvalue and normalized eigenvector of $\rho $ (respectively, $\sigma $)%
$.$ Clearly, $q_{k}=\sum_{k^{\prime }}D_{kk^{\prime }}p_{k^{\prime
}}$ where $D_{kk^{\prime }}\equiv {\mathrm{Tr}}\left( \left\vert
f_{k}\right\rangle \left\langle f_{k}\right\vert
\mathcal{E}(\left\vert e_{k^{\prime }}\right\rangle \left\langle
e_{k^{\prime }}\right\vert )\right) .$ From the
fact that $\mathcal{E}$ is trace-preserving one infers that $%
\sum_{k}D_{kk^{\prime }}=1$ while from the fact that $\mathcal{E}$
is unital one infers that $\sum_{k^{\prime }}D_{kk^{\prime }}=1$.
It follows that $ D=(D_{kk^{\prime}})$ is bistochastic. Thus, by
the Hardy-Littlewood-Polya theorem, we have $\mathbf{r} \prec
\mathbf{s.}$ QED

\strut

Heuristically, this says that unital channels can only
\emph{increase} the impurity or \textquotedblleft
mixedness\textquotedblright\ of quantum states.

\begin{corollary}
\label{corollaryofuhlmann} For a unital operation $\mathcal{E}$
acting on a projector $P$, we have
$\mathrm{rank}(\mathcal{E}(P))\ge \mathrm{rank}(P)$.
\end{corollary}

This is a simple consequence of the majorisation relation given by
Eq.~(\ref{uhlmanneqn}).

\begin{corollary}
\label{supportlemma} Let $\mathcal{E}$ be a unital operation and
$P$ a projection. If $\mathrm{rank}(\mathcal{E}(P))=
\mathrm{rank}(P)$, that is,
if the inequality of Corollary \ref{corollaryofuhlmann} is saturated, then $%
\mathcal{E}(P)$ is a projection.
\end{corollary}

\noindent\textit{Proof.} A projection $P$ has a uniform spectrum,
and the only spectrum that is majorised by the uniform spectrum
while having the
same number of non-zero elements is the uniform spectrum. Given that $%
\mathcal{E}$ is trace-preserving, it follows that $\mathcal{E}(P)$
is a projection. QED

\begin{lemma}
\label{fpofEisfpofEdag} For a unital operation $\mathcal{E}$
acting on a projector $P$, we have $\mathcal{E}(P)=P$ if and only
if $\mathcal{E}^{\dag }(P)=P.$
\end{lemma}

\textit{Proof. \ }By the fixed point theorem for unital operations
\cite{Kri03a}, $P$ is in the fixed point set of $\mathcal{E}$ if
and only if it commutes with all Kraus operators of $\mathcal{E}$.
The lemma then follows from the fact that the Kraus operators of $\mathcal{E}%
^{\dag }$ are the adjoints of those of $\mathcal{E}$ and that both
maps are unital. QED


We are now in a position to prove Lemma \ref{newlemma}.


\textit{Proof of Lemma \ref{newlemma}.} For $(ii)\Rightarrow (i)$,
we assume that $
\mathrm{rank}(\mathcal{E}(P_{AB}))=\mathrm{rank}(P_{AB})$. It
follows from the fact that $\mathcal{E}$ is unital and Corollary
\ref{supportlemma}, that
$\mathcal{E}(P_{AB})$ is a projection. \ A projection of the same rank as $%
P_{AB}$ is obtained by a unitary operation $\mathcal{U}$ acting on
$P_{AB}$,
so that $\mathcal{E}(P_{AB})=\mathcal{U}(P_{AB}).$ We infer that $\mathcal{U}%
^{\dag }\circ \mathcal{E}(P_{AB})=P_{AB}.$ By
Lemma~\ref{fpofEisfpofEdag}, it follows that $\mathcal{E}^{\dag
}\circ
\mathcal{U}(P_{AB})=P_{AB}$. Consequently we have $\mathcal{E}^{\dag }\circ \mathcal{E}(P_{AB})=\mathcal{E}%
^{\dag }\circ \mathcal{U}\circ \mathcal{U}^{\dag }\circ \mathcal{E}%
(P_{AB})=P_{AB},$ and {\it a fortiori} condition $(i)$ holds.

For $(i)\Rightarrow (ii)$, we assume  that \strut
\textrm{supp}$(\mathcal{E}^{\dag }\circ
\mathcal{E}(P_{AB}))\subseteq $ \textrm{supp}$(P_{AB})$. It
follows that \textrm{rank}$(\mathcal{E}^{\dag }\circ
\mathcal{E}(P_{AB}))\leq $ \textrm{rank}$(P_{AB})$. Thus, given
that $\mathcal{E}^{\dag }\circ \mathcal{E} $ is a unital
operation, by Corollary \ref{corollaryofuhlmann}, we deduce that
in fact \textrm{rank}$(\mathcal{E}%
^{\dag }\circ \mathcal{E}(P_{AB}))=$ \textrm{rank}$(P_{AB})$. \ By Corollary %
\ref{supportlemma}, it follows that $\mathcal{E}^{\dag }\circ \mathcal{E}%
(P_{AB})$ is a projection. The only projection with the same rank
and support as $P_{AB}$ is $P_{AB},$ therefore $\mathcal{E}^{\dag
}\circ \mathcal{E}(P_{AB})=P_{AB}.$ \ Given that
$\mathcal{E}^{\dag }$ is unital,
it follows from Corollary \ref{corollaryofuhlmann} that $\mathrm{rank%
}(\mathcal{E}^{\dag }\circ \mathcal{E}(P_{AB}))\geq \mathrm{rank}(\mathcal{E}%
(P_{AB}))$. \ Given that $\mathcal{E}^{\dag }\circ \mathcal{E}%
(P_{AB})=P_{AB},$ we deduce that $\mathrm{rank}(P_{AB})\geq \mathrm{rank}(%
\mathcal{E}(P_{AB})).$ \ However, from Corollary
\ref{corollaryofuhlmann} and the unitalness of $\mathcal{E},$ we
infer the opposite inequality
$\mathrm{rank}(\mathcal{E}(P_{AB}))\geq \mathrm{rank}(P_{AB}),$
so that we must have equality, $\mathrm{rank}(\mathcal{E}(P_{AB}))=\mathrm{%
rank}(P_{AB}).$ QED

\strut

Notice from this proof that the two conditions of
Lemma~\ref{newlemma} are equivalent to the seemingly stronger
statement $\mathcal{E}^\dagger\circ\mathcal{E}(P_{AB}) = P_{AB}$.


\begin{thebibliography}{30}
\expandafter\ifx\csname
natexlab\endcsname\relax\def\natexlab#1{#1}\fi
\expandafter\ifx\csname bibnamefont\endcsname\relax
  \def\bibnamefont#1{#1}\fi
\expandafter\ifx\csname bibfnamefont\endcsname\relax
  \def\bibfnamefont#1{#1}\fi
\expandafter\ifx\csname citenamefont\endcsname\relax
  \def\citenamefont#1{#1}\fi
\expandafter\ifx\csname url\endcsname\relax
  \def\url#1{\texttt{#1}}\fi
\expandafter\ifx\csname
urlprefix\endcsname\relax\def\urlprefix{URL }\fi
\providecommand{\bibinfo}[2]{#2}
\providecommand{\eprint}[2][]{\url{#2}}

\bibitem{VKL01} \bibinfo{author}{\bibfnamefont{L.}~\bibnamefont{Viola}}, %
\bibinfo{author}{\bibfnamefont{E.}~\bibnamefont{Knill}} and %
\bibinfo{author}{\bibfnamefont{R.}~\bibnamefont{Laflamme}}, %
\bibinfo{journal}{J. Phys. {A}} \textbf{\bibinfo{volume}{34}}, %
\bibinfo{pages}{7067} (\bibinfo{year}{2001}).

\bibitem{Kni06} \bibinfo{author}{\bibfnamefont{E.} \bibnamefont{Knill}}, %
\bibinfo{journal}{arxiv.org/quant-ph/0603252}.

\bibitem{KLP05} \bibinfo{author}{\bibfnamefont{D.~W.}~\bibnamefont{Kribs}}, %
\bibinfo{author}{\bibfnamefont{R.}~\bibnamefont{Laflamme}} and %
\bibinfo{author}{\bibfnamefont{D.}~\bibnamefont{Poulin}}, %
\bibinfo{journal}{Phys. Rev. Lett.},\textbf{\bibinfo{volume}{94}}, %
\bibinfo{pages}{180501} (\bibinfo{year}{2005}).

\bibitem{KLPL05} \bibinfo{author}{\bibfnamefont{D.~W.}~\bibnamefont{Kribs}}, %
\bibinfo{author}{\bibfnamefont{R.}~\bibnamefont{Laflamme}}, %
\bibinfo{author}{\bibfnamefont{D.}~\bibnamefont{Poulin}} and %
\bibinfo{author}{\bibfnamefont{M.}~\bibnamefont{Lesosky}}, %
\bibinfo{journal}{Quantum Inf. \& Comp.} \textbf{\bibinfo{volume}{6}} (%
\bibinfo{year}{2006}), \bibinfo{pages}{382-399}.

\bibitem[Shor(1995)]{Sho95a}
\bibinfo{author}{\bibfnamefont{P.~W.}
\bibnamefont{Shor}}, \bibinfo{journal}{Phys. Rev. A} \textbf{%
\bibinfo{volume}{52}}, \bibinfo{pages}{R2493} (\bibinfo{year}{1995}).

\bibitem[Steane(1996)]{Ste96a}
\bibinfo{author}{\bibfnamefont{A.~M.}
\bibnamefont{Steane}}, \bibinfo{journal}{Phys. Rev. Lett.} \textbf{%
\bibinfo{volume}{77}}, \bibinfo{pages}{793} (\bibinfo{year}{1996}).

\bibitem[Gottesman(1996)]{Got96}
\bibinfo{author}{\bibfnamefont{D.}
\bibnamefont{Gottesman}}, \bibinfo{journal}{Phys. Rev. A} \textbf{%
\bibinfo{volume}{54}} \bibinfo{pages}{1862} (\bibinfo{year}{1996}).

\bibitem[Bennett et~al.(1996)Bennett, DiVincenzo, Smolin, and Wootters]%
{BDSW96a} \bibinfo{author}{\bibfnamefont{C.~H.} \bibnamefont{Bennett}}, %
\bibinfo{author}{\bibfnamefont{D.~P.} \bibnamefont{DiVincenzo}}, %
\bibinfo{author}{\bibfnamefont{J.~A.} \bibnamefont{Smolin}}, and
\bibinfo{author}{\bibfnamefont{W.~K.}
  \bibnamefont{Wootters}}, \bibinfo{journal}{Phys. Rev. A} \textbf{%
\bibinfo{volume}{54}}, \bibinfo{pages}{3824} (\bibinfo{year}{1996}).

\bibitem[Knill and Laflamme(1997)]{KL97a} \bibinfo{author}{%
\bibfnamefont{E.}~\bibnamefont{Knill}} and \bibinfo{author}{%
\bibfnamefont{R.}~\bibnamefont{Laflamme}}, \bibinfo{journal}{Phys. Rev. {A}}
\textbf{\bibinfo{volume}{55}}, \bibinfo{pages}{900}
(\bibinfo{year}{1997}).

\bibitem[Palma et~al.(1996)]{PSE96} \bibinfo{author}{\bibfnamefont{G.M.}~%
\bibnamefont{Palma}}, \bibinfo{author}{\bibfnamefont{K.-A.}~%
\bibnamefont{Suominen}} and \bibinfo{author}{\bibfnamefont{A.}~%
\bibnamefont{Ekert}}, \bibinfo{journal}{Proc. Royal Soc. A} \textbf{%
\bibinfo{volume}{452}}, \bibinfo{pages}{567} (\bibinfo{year}{1996}).

\bibitem[Duan and Guo(1997)]{DG97c}
\bibinfo{author}{\bibfnamefont{L.-M.}
\bibnamefont{Duan}} and
\bibinfo{author}{\bibfnamefont{G.-C.}
\bibnamefont{Guo}}, \bibinfo{journal}{Phys. Rev. Lett.} \textbf{%
\bibinfo{volume}{79}}, \bibinfo{pages}{1953} (\bibinfo{year}{1997}).

\bibitem[Zanardi and Rasetti(1997)]{ZR97c} \bibinfo{author}{%
\bibfnamefont{P.}~\bibnamefont{Zanardi}} and \bibinfo{author}{%
\bibfnamefont{M.}~\bibnamefont{Rasetti}},
\bibinfo{journal}{Phys. Rev.
Lett.} \textbf{\bibinfo{volume}{79}}, \bibinfo{pages}{3306} (%
\bibinfo{year}{1997}).

\bibitem[Lidar et~al.(1998)Lidar, Chuang, and Whaley]{LCW98a} %
\bibinfo{author}{\bibfnamefont{D.A.}~\bibnamefont{Lidar}}, %
\bibinfo{author}{\bibfnamefont{I.L.}~\bibnamefont{Chuang}}, and %
\bibinfo{author}{\bibfnamefont{K.B.}~\bibnamefont{Whaley}}, %
\bibinfo{journal}{Phys. Rev. Lett.} \textbf{\bibinfo{volume}{81}}, %
\bibinfo{pages}{2594} (\bibinfo{year}{1998}).

\bibitem[Knill et~al.(2000)Knill, Laflamme, and Viola]{KLV00a} %
\bibinfo{author}{\bibfnamefont{E.}~\bibnamefont{Knill}}, \bibinfo{author}{%
\bibfnamefont{R.}~\bibnamefont{Laflamme}}, and \bibinfo{author}{%
\bibfnamefont{L.}~\bibnamefont{Viola}}, \bibinfo{journal}{Phys. Rev. Lett.}
\textbf{\bibinfo{volume}{84}}, \bibinfo{pages}{2525}
(\bibinfo{year}{2000}).

\bibitem[Zanardi(2001)]{Zan01b} \bibinfo{author}{\bibfnamefont{P.}~%
\bibnamefont{Zanardi}}, \bibinfo{journal}{Phys. Rev. A} \textbf{%
\bibinfo{volume}{63}}, \bibinfo{pages}{12301} (\bibinfo{year}{2001}).

\bibitem[Kempe et~al.(2001)Kempe, Bacon, Lidar, and Whaley]{KBLW01a} %
\bibinfo{author}{\bibfnamefont{J.}~\bibnamefont{Kempe}}, \bibinfo{author}{%
\bibfnamefont{D.}~\bibnamefont{Bacon}}, \bibinfo{author}{%
\bibfnamefont{D.~A.} \bibnamefont{Lidar}}, and \bibinfo{author}{%
\bibfnamefont{K.~B.} \bibnamefont{Whaley}}, \bibinfo{journal}{Phys. Rev. A}
\textbf{\bibinfo{volume}{63}}, \bibinfo{pages}{42307}
(\bibinfo{year}{2001}).

\bibitem{Bac05} \bibinfo{author}{\bibfnamefont{D.} \bibnamefont{Bacon}}, %
\bibinfo{journal}{Phys. Rev. A} \textbf{\bibinfo{volume}{73}}, %
\bibinfo{pages}{012340} (\bibinfo{year}{2006}).

\bibitem{Pou05} \bibinfo{author}{\bibfnamefont{D.}~\bibnamefont{Poulin}}, %
\bibinfo{journal}{Phys. Rev. Lett.} \textbf{\bibinfo{volume}{95}}, %
\bibinfo{pages}{230504} (\bibinfo{year}{2005}).

\bibitem{KlaSar06} \bibinfo{author}{\bibfnamefont{A.}~%
\bibnamefont{Klappenecker}} and \bibinfo{author}{\bibfnamefont{P.}~%
\bibnamefont{Kiran Sarvepalli}}, %
\bibinfo{journal}{arxiv.org/quant-ph/0604161}.

\bibitem{CK06} \bibinfo{author}{\bibfnamefont{M.~D.}~\bibnamefont{Choi}} and %
\bibinfo{author}{\bibfnamefont{D.~W.}~\bibnamefont{Kribs}}, %
\bibinfo{journal}{Phys. Rev. Lett.} \textbf{\bibinfo{volume}{96}}, %
\bibinfo{pages}{050501} (\bibinfo{year}{2006}).

\bibitem{NC00} \bibinfo{author}{\bibfnamefont{M.~A.}~\bibnamefont{Nielsen}},
and \bibinfo{author}{\bibfnamefont{I.}~\bibnamefont{Chuang}}, %
   \bibinfo{journal}{{\it Quantum computation and quantum information},}
\bibinfo{journal}{Cambridge University Press} (\bibinfo{year}{2000}).

\bibitem{NaySen06} \bibinfo{author}{\bibfnamefont{A.}~\bibnamefont{Nayak}},
and \bibinfo{author}{\bibfnamefont{P.}~\bibnamefont{Sen}}, %
\bibinfo{journal}{Quantum Inf. \& Comp., to appear}.

\bibitem{NP05} \bibinfo{author}{\bibfnamefont{M.~A.}~\bibnamefont{Nielsen}}
and \bibinfo{author}{\bibfnamefont{D.}~\bibnamefont{Poulin}}, %
\bibinfo{journal}{arxiv.org/quant-ph/0506069}.

\bibitem{NCSB97} \bibinfo{author}{\bibfnamefont{M.~A.}~\bibnamefont{Nielsen}}%
, \bibinfo{author}{\bibfnamefont{C.~M.}~\bibnamefont{Caves}}, %
\bibinfo{author}{\bibfnamefont{B.} \bibnamefont{Schumacher}}, and %
\bibinfo{author}{\bibfnamefont{H.} \bibnamefont{Barnum}}, %
\bibinfo{journal}{Proc. Royal Soc. A} \textbf{\bibinfo{volume}{454}}, %
\bibinfo{pages}{277} (\bibinfo{year}{1997}).

\bibitem[Kribs(2003)]{Kri03a}
\bibinfo{author}{\bibfnamefont{D.~W.}
\bibnamefont{Kribs}}, \bibinfo{journal}{Proc. Edin. Math. Soc.} \textbf{%
\bibinfo{volume}{46}}, \bibinfo{pages}{421}(\bibinfo{year}{2003}).

\bibitem[Holbrook, et al.(2004)Holbrook, Kribs, and Laflamme]{HKL04} %
\bibinfo{author}{\bibfnamefont{J.~A.}~\bibnamefont{Holbrook}}, %
\bibinfo{author}{\bibfnamefont{D.~W.}~\bibnamefont{Kribs}}, and %
\bibinfo{author}{\bibfnamefont{R.}~\bibnamefont{Laflamme}}, %
\bibinfo{journal}{Quantum Inf. Proc.} \textbf{\bibinfo{volume}{2}}, %
\bibinfo{pages}{381} (\bibinfo{year}{2004}).

\bibitem{CKZ05b} \bibinfo{author}{\bibfnamefont{M.~D.}~\bibnamefont{Choi}}, %
\bibinfo{author}{\bibfnamefont{D.~W.}~\bibnamefont{Kribs}}, and %
\bibinfo{author}{\bibfnamefont{K.}~\bibnamefont{{\.Z}yczkowski}},
\bibinfo{journal}{Rep. Math. Phys.,}
\textbf{\bibinfo{volume}{58}}, %
\bibinfo{pages}{77} (\bibinfo{year}{2006}).

\bibitem{Blu05}
\bibinfo{author}{\bibfnamefont{R.}
\bibnamefont{Blume-Kohout}}, \bibinfo{journal}{PhD thesis, Berkeley} (%
\bibinfo{year}{2005}).

\bibitem{HLP34}
\bibinfo{author}{\bibfnamefont{G.~H.}~\bibnamefont{Hardy}},
{\bibfnamefont{J.~E.}~\bibnamefont{Littlewood}},
and \bibinfo{author}{\bibfnamefont{G.}~\bibnamefont{Polya}}, %
   \bibinfo{journal}{{\it Inequalities},}
\bibinfo{journal}{Cambridge University Press} (\bibinfo{year}{1934}).

\bibitem{Bha97} \bibinfo{author}{\bibfnamefont{R.}~\bibnamefont{Bhatia}},
   \bibinfo{journal}{{\it Matrix Analysis},}
\bibinfo{journal}{Springer, New York,} (\bibinfo{year}{1997}).

\bibitem{Uhl77}
\bibinfo{author}{\bibfnamefont{A.}
\bibnamefont{Uhlmann}}, \bibinfo{journal}{Commun. Math. Phys.} \textbf{%
\bibinfo{volume}{54}}, \bibinfo{pages}{21}(\bibinfo{year}{1977}).

\bibitem{BZ06} \bibinfo{author}{\bibfnamefont{I.}~\bibnamefont{Bengtsson}},
and \bibinfo{author}{\bibfnamefont{K.}~\bibnamefont{{\.Z}yczkowski}}, %
   \bibinfo{journal}{{\it Geometry of quantum states},}
\bibinfo{journal}{Cambridge University Press} (\bibinfo{year}{2006}).


\end{thebibliography}
\end{document}